\documentclass[useams,usenatbib]{mn2e}

\usepackage{natbib}
\usepackage{amsmath}
\usepackage{amssymb}
\usepackage{graphicx}
\usepackage{url}
\usepackage{hyperref}
\newcommand{\spider}{{\sc Spider}}
\newcommand{\ebex}{{\sc EBEX}}
\newcommand{\spt}{{\sc SPT}}
\newcommand{\archeops}{{\sc Archeops}}

\newcommand{\WMAP}{{\sc WMAP}}
\newcommand{\IRAS}{{\sc IRAS}}

\newcommand{\healpix}{{\sc healpix}}

\newcommand{\dust}{\mathrm{dust}}

\newcommand{\fds}{\mathrm{FDS}}

\newcommand{\sech}{\mathrm{\,sech\,}}
\newcommand{\nn}{\nonumber}
\newcommand{\ud}{{\mathrm{d}}}
\newcommand{\vect}[1]{\ensuremath{\boldsymbol{#1}}}

\newcommand{\sscale}{{\it ss}}
\newcommand{\lscale}{{\it ls}}
\newcommand{\abs}{{\sc ABS}}
\newcommand{\polar}{{\sc POLAR}}
\newcommand{\polarbear}{{\sc POLARBEAR}}
\newcommand{\brain}{{\sc BRAIN}}
\newcommand{\act}{{\sc ACT}}
\newcommand{\piper}{{\sc PIPER}}
\newcommand{\planck}{{\sc Planck}}

\providecommand{\adsurl}[1]{\href{#1}{ADS}}

\title[Foreground Emission From Interstellar Dust]{A Model For Polarised Microwave Foreground Emission From Interstellar Dust}
\author[D.~T.~O'Dea, C.~N.~Clark, C.~R.~Contaldi and C.~J.~MacTavish]{D.~T.~O'Dea$^{1}$,  C.~N.~Clark$^{1}$\thanks{E-mail:
caroline.clark05@imperial.ac.uk},  C.~R.~Contaldi$^{1}$ and C.~J.~MacTavish$^{2}$\\
$^{1}$Theoretical Physics, Blackett Laboratory, Imperial
  College, London, UK\\
$^{2}$Kavli Institute for Cosmology, University of Cambridge, Cambridge, UK}
\begin{document}


\pagerange{\pageref{firstpage}--\pageref{lastpage}} \pubyear{2011}

\maketitle

\label{firstpage}

\begin{abstract}
  The upcoming generation of cosmic microwave background (CMB)
  experiments face a major challenge in detecting the weak cosmic
  $B$-mode signature predicted as a product of primordial
  gravitational waves. To achieve the required sensitivity these
  experiments must have impressive control of systematic effects and
  detailed understanding of the foreground emission that will
  influence the signal. In this paper, we present templates of the
  intensity and polarisation of emission from one of the main Galactic
  foregrounds, interstellar dust. These are produced using a model
  which includes a 3D description of the Galactic magnetic field,
  examining both large and small scales. We also include in the model
  the details of the dust density, grain alignment and the intrinsic
  polarisation of the emission from an individual grain. We present
  here Stokes parameter template maps at $150$ GHz and provide an
  on-line
  repository\footnote{\url{http://www.imperial.ac.uk/people/c.contaldi/fgpol}}
  for these and additional maps at frequencies that will be targeted
  by upcoming experiments such as \ebex, \spider\ and \spt pol.
\end{abstract}


\section{Introduction}

The next round of cosmic microwave background (CMB) experiments are
all targeting measurements of CMB polarisation.  The CMB
  polarisation field can be decomposed into curl-free $E$-modes and
  curl-like $B$-modes (see for example \cite{1997PhRvD..55.7368K}),
  however to date only $E$-modes have been observed
  \citep{DASI,CBIpol,2007ApJ...660..976S, CAPMAP,2008ApJ...684..771B,
    DASI3yr,BoomerangEE, 2006ApJ...647..833P,2007ApJ...665...55W,
    2007ApJS..170..335P,2009ApJ...705..978B, 2010ApJ...711.1123C,
    2010arXiv1012.3191Q}. Current work centres on the search
  for the tiny amplitude $B$-modes since a key prediction of inflation
  is the generation of a stochastic background of gravitational waves,
  which on large angular scales are the only contribution to the CMB
  $B$-mode component. Experiments involved in this search include
  \ebex\ \citep{2010SPIE.7741E..37R}, \spider\
  \citep{2010SPIE.7741E..46F}, \spt pol~\citep{2009AIPC.1185..511M},
  \piper\ \citep{2011AAS...21823301L}, \abs\
  \citep{2010arXiv1008.3915E}, \act pol \citep{2010SPIE.7741E..51N},
  \polar\ \citep{1998ApJ...495..580K}, \polarbear\
  \citep{2010arXiv1011.0763T} and \brain\
  \citep{2008arXiv0805.4527C}. We
  focus on providing templates at frequencies being targeted by \ebex
  , \spider\ and \spt pol. Galactic foreground emission is expected
to contribute significantly to the polarised microwave emission across
the sky, and may well dominate the CMB gravitational-wave signal at
all frequencies.  In order to achieve their scientific goals,
forthcoming CMB polarisation experiments require in-depth knowledge of
this polarised Galactic foreground emission. The \planck\ mission
  \citep{2006astro.ph..4069T} will provide maps of the polarisation of
  interstellar dust, allowing tests of the structure of the Galactic
  magnetic field. It should also provide an insight into grain alignment
  mechanisms.

Of utmost importance will be the accurate separation of foreground
emission from the CMB signal. Component separation has been
  considered in, for example, \citet{1994ApJ...424....1B,
    2006ApJ...641..665E, 2007ApJ...665..355K, 2009MNRAS.392..216S}. In
order to test and assess methods for separating out the contribution
realistic foreground templates will be required. Since however,
  few data exist at this time at the observing frequencies in which
the upcoming experiments are operating, one must resort to modeling of
foreground emission by extrapolating the information from existing
data.  Furthermore, experiments which observe portions of the sky
close to the Galactic plane will find the foreground emission is very
bright in comparison to the CMB. The presence of this bright emission
in the data may affect the performance of the observation and the
analysis strategy an experiment uses. Another important role of
foreground modeling, therefore, is informing the planning and proposal
stage of any experiment.

Unfortunately, these foregrounds are poorly constrained by current
data and poorly understood, particularly above around $90$\,GHz, where
the CMB emission is strongest and where many new CMB experiments will
operate.  At these frequencies, the foreground emission is expected to
be dominated by thermal emission from interstellar dust. A review
  of the basic physical processes whereby aligned dust grains generate
  polarisation is given in \citet{2003rrda.book..255L}.
Such emission is known to be polarised both through direct
measurements \citep{2005A&A...444..327P, 2007ApJ...665..355K,
  2004A&A...424..571B,Bierman:2011hq} and through observations of the
polarisation of starlight \citep{1996ApJ...462..316H,
  2002ApJ...564..762F}.

 This polarisation
arises due to the presence of a magnetic field in the Galaxy. The dust
grains are generally non-spherical, and preferentially emit radiation
polarised along their longest axis. Mechanisms exist which align these
grains with this axis perpendicular to the Galactic magnetic field,
leading to net linear polarisation.

\begin{figure*}
\begin{center}\begin{tabular}{c}
    \makebox[3in][c]{\includegraphics[height=7cm, width=7cm,angle=0,trim=0cm 0cm 0cm 0cm,clip]{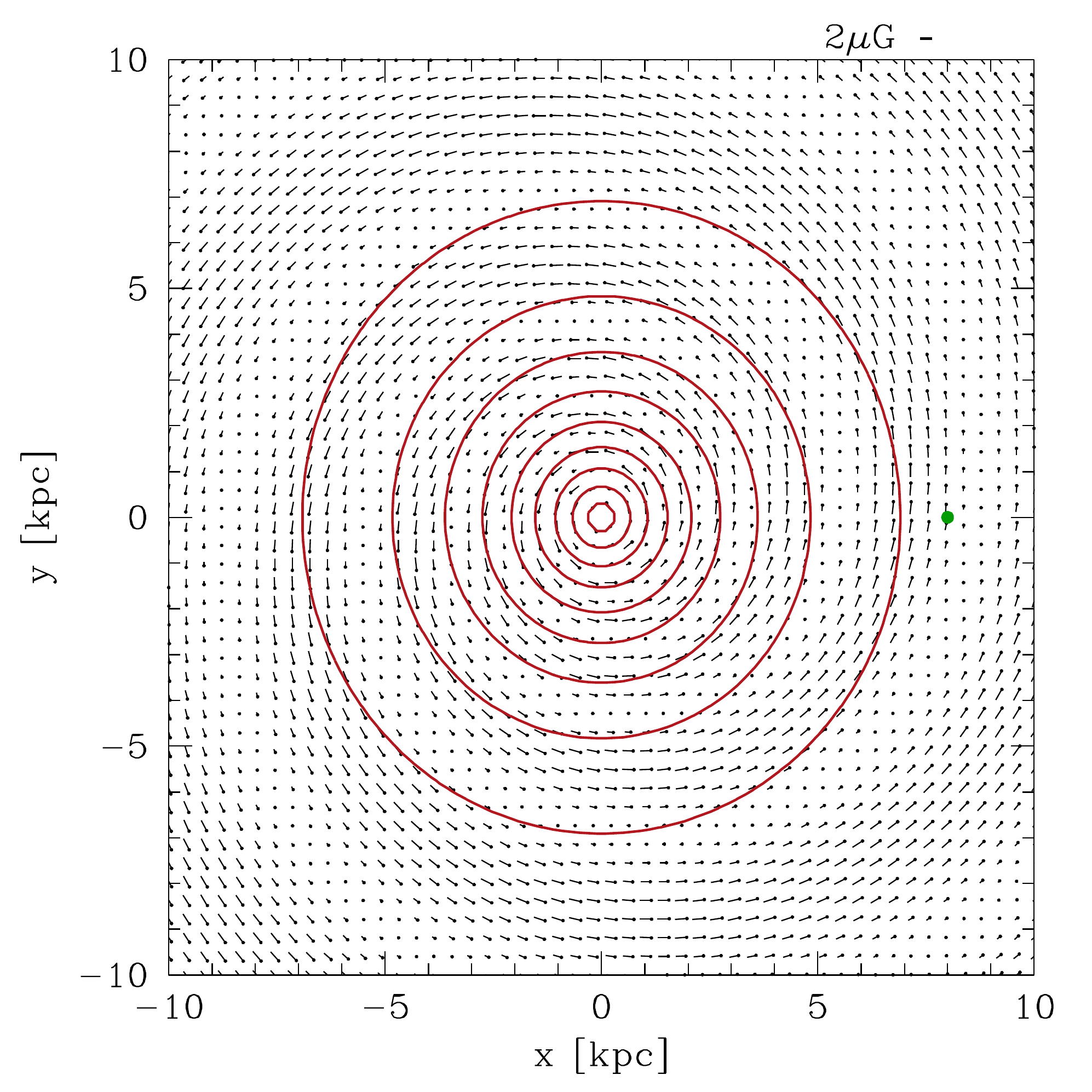}\includegraphics[height=7cm, width=7cm,angle=0,trim=0cm 0cm 0cm 0cm,clip]{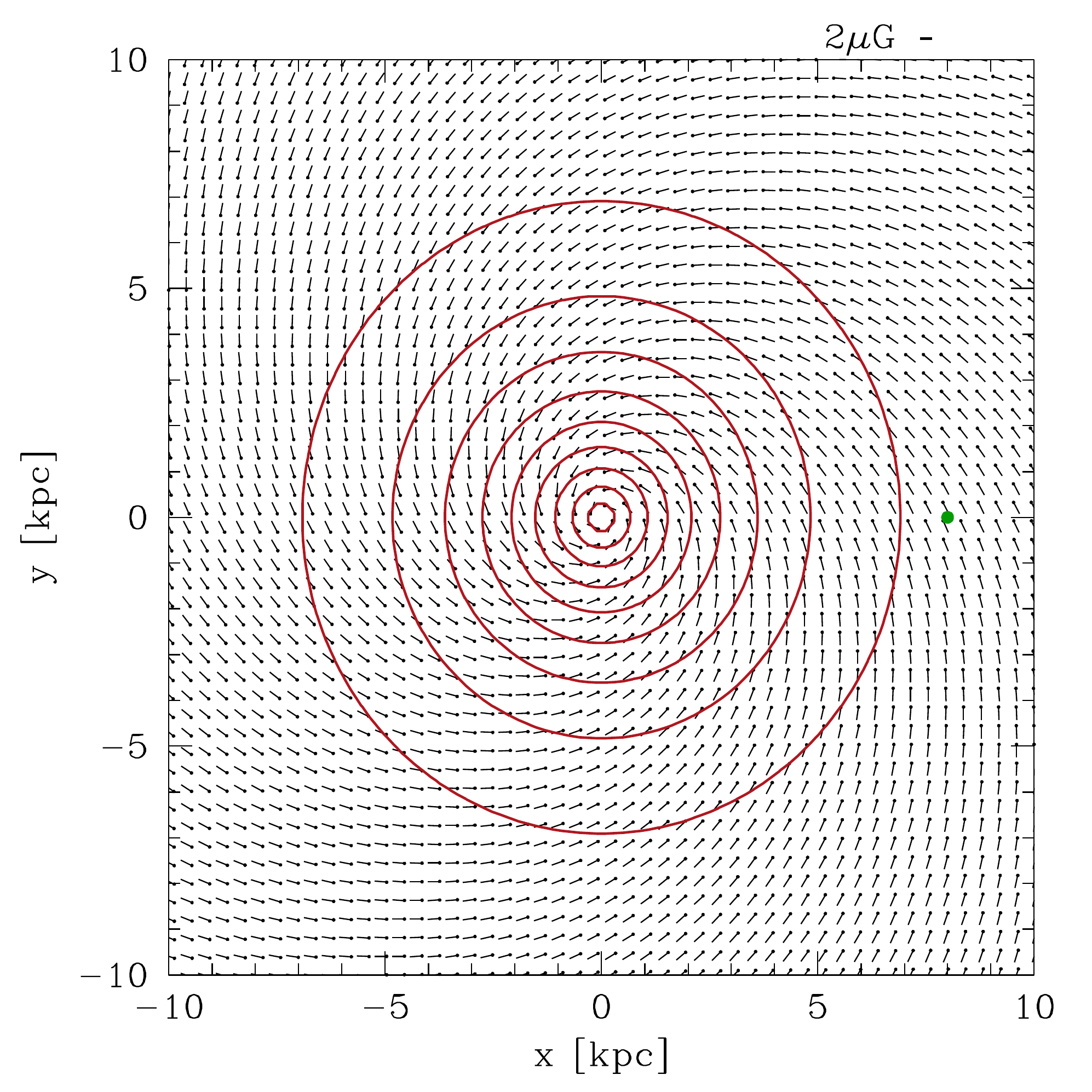}} \\
\end{tabular}
\caption[]{Cartesian projection of the BSS (left panel) and LSA (right
  panel) large-scale magnetic field models, showing a slice through
  the Galactic plane observed along the positive $z$-axis. The filled
  circle represents the position of the Sun in relation to the
  Galactic centre. The alignment and magnitude of the magnetic field
  are shown as headed ticks with the $B_0=2$ $\mu$G scale represented
  at the top of each panel. The red solid lines are the density
  contours, in steps of 0.1, of the dust density model $n_d(r,z)$ (see
  Section~\ref{sec:dustdens}) used in the line-of-sight integration.
  The dust density is normalised to 1 at the Galactic centre.}
\label{fig:lsmodel}
\end{center}
\end{figure*}

Polarised foregrounds also include polarised emission from synchrotron.
Synchrotron emission, generated by the gyration of cosmic ray
electrons in the Galactic magnetic field, is intrinsically polarised
and constitutes the main polarised foreground at lower frequencies
\citep{2007ApJS..170..335P}.  However, emission from thermal dust
dominates synchrotron emission at the frequencies considered in this
paper and therefore we concentrate on modeling emission from only thermal
dust. Foreground radiation also includes free-free and 
spinning dust emission, however we assume both of these signals to be
unpolarised and so do not consider them further here. Evidence
  for this has emerged recently, with \citet{2011arXiv1108.0205M} showing that free-free
  emission is unpolarised, setting an upper limit on the free-free polarisation fraction of 3.4\% at the $2\sigma$ level. They also show that spinning dust emission has a low polarisation fraction. Upper limits on the polarisation fraction of spinning dust emission in molecular clouds have been obtained by \citet{2011arXiv1108.0308D} and \citet{2011ApJ...729...25L}, who find low levels of polarisation. If there is a similar level at higher Galactic latitude, then this foreground is unimportant in terms of component separation. 

The model in this paper was introduced by \cite{Danielthesis:2009} and
first applied in~\citet{2011arXiv1102.0559O} for the purpose of
studying the impact of polarised foregrounds on Spider's ability to
detect $B$-mode polarisation.  Here we give a more detailed
explanation of the model and present a number of full-sky template
maps at various frequencies.  \archeops\ 
\citep{2004A&A...424..571B} and {\sc BICEP} \citep{Bierman:2011hq}
have made the highest signal--to--noise maps of the dust polarisation
at low Galactic latitudes and have examined the properties of the
polarisation fraction and angle. These results cannot be relied upon
to calibrate large scale models of the polarisation at higher
latitudes since the polarisation properties near the Galactic plane
will depend on complex structure that is not included in models such
as the one presented in this work. Here, both polarisation amplitude
and angle are modeled internally and our templates are scaled such
that the polarisation fraction corresponds to a nominal value when
averaged over the maps with the Galaxy masked out. The \archeops\ 
and {\sc BICEP} maps have not been made public and a quantitative
comparison of our templates with these observations is not
possible. \citet{AA..526..A145..2011} have developed a similar model
of both the polarisation of thermal dust and synchrotron radiation and
compared with the \WMAP\ K-band and \archeops\  353GHz data, however
they have not released any templates based on their model.

This paper is organized as follows.  We begin in
Section~\ref{sec:total} by describing the model of the dust total
intensity.  We focus on modeling the Galactic magnetic field (which is
made up of a large-scale field from the Galactic disk and a
small-scale field due to turbulence in the interstellar plasma) in
Section~\ref{sec:galacticfield}.  We include in
Section~\ref{sec:dustdens} a description of the large-scale dust
density, as well as details of the treatment of dust grain alignment
and the intrinsic polarisation of the emission from an individual
grain. In Section~\ref{sec:stokes} we show how these fields are
modeled in three-dimensions and are used to perform a line-of-sight
integral to the centre of each pixel to form the final Stokes
parameter maps.  In Section~\ref{sec:scales} we discuss the effective
resolution of the maps produced using our model and the relevant
physical scales introduced by various effects. In
Section~\ref{sec:maps} we describe the final template maps and
summarise our conclusions in Section~\ref{sec:conclude}.

\section{Foreground Dust Model}
\label{sec:skymodel}

\subsection{Dust Total Intensity}
\label{sec:total}

Although few data are available regarding the polarised emission
from dust, the same is not true of its total intensity. In particular,
the \IRAS\ satellite observed this emission across the sky at $100
\mu$m and $240 \mu$m, close to the peak in the dust emission. By
constraining physically-motivated extrapolations of these observations
using further data, \citet[hereafter FDS]{1999ApJ...524..867F}
provided models of the emission at microwave wavelengths. At $94$\,GHz, 
these models have been shown to agree well in terms of
morphology with the \WMAP\ observations with some minor structural
differences on the Galactic Plane \citep{2009ApJS..180..265G}.
However, there are indications that in terms of amplitude the \WMAP\
dust template fit coefficients differ by about 30\%.
\citet{2003ApJS..148...97B} suggest that this is possibly due to the
degeneracy that exists between the strongly correlated dust and
synchrotron emission components in the simultaneous fit of their
externally derived template maps to \WMAP\ data.

In the higher frequency bands relevant to experiments
observing above $\sim90$ GHz, data are more limited but
  agree well with the FDS predictions \citep{2010ApJ...722.1057C,
  2010ApJ...713..959V}. We will use this model (to be precise model
number eight in FDS) to trace the total intensity of the dust
emission. We exploit the full, 6.1 arcminutes, resolution of the
\IRAS\ data by pixelising the dust intensity on \healpix\
\citep{2005ApJ...622..759G} maps of $N_{\rm side}$=1024.

\subsection{Galactic Magnetic Field}
\label{sec:galacticfield}

The
degree and direction of polarisation of the dust emission are highly dependent on the
Galactic magnetic field. As the observed polarisation is the sum of
many independent regions along the line-of-sight, it is sensitive to
the three-dimensional structure of this magnetic field. Therefore, to
proceed we first consider a three-dimensional model of the Galactic
magnetic field (and the other necessary Galactic constituents and
physics) and then set the polarisation degree and direction through
the appropriate line-of-sight integrals.

Away from the Galactic center, the Galactic magnetic field is usually
considered to have two near-independent components: a large-scale
coherent field associated with the Galactic disk, and a small-scale
field arising from turbulence in the interstellar plasma sourced by
astrophysical events such as supernovae and stellar winds. The most
informative probes of these fields are Faraday rotation measures of
pulsars and extra-Galactic radio sources
\citep{2006ApJS..167..230H,2006ApJ...642..868H}. Whilst there is
general agreement that the large-scale field follows a spiral pattern,
its detailed structure is still uncertain.

When considering areas of sky at high Galactic latitudes, this
uncertainty is unimportant as the dust is concentrated in a thin disk
about the Galactic plane, and so we only see emission within around
$1$\,kpc or so of the Sun, a region in which the large-scale field is
reasonably well characterised. However, experiments which will target a
large fraction of the sky, possibly including part of the Galactic
plane, will require a model of the large-scale field structure in the
plane.

Attempts have been made to constrain the properties of the magnetic
field using CMB polarisation measurements.  \citet{Jaffe:2009hh} use
an Monte Carlo Markov Chain (MCMC) approach to test components of a 2D
Galactic field model using rotation measures and \WMAP\ data in the
plane of the Galaxy. \citet{2009JCAP...07..021J} use rotation measures and
\WMAP\ 5 year data to fit for parameters in common 3D models for the
Galactic magnetic field. We choose two of the most popular forms for
the Galactic magnetic field and provide templates for both of these
models.

\subsubsection{Large-Scale Magnetic Field}

One popular candidate is the Bi-Symmetric Spiral (BSS)
\citep{1994A&A...288..759H, 2008A&A...477..573S} which can be written
as
\begin{align}
  B_{\rho} &= -B_0 \cos{\left(\Phi + \psi
      \ln{\frac{\rho}{\rho_0}}\right)} \sin{p} \cos{\chi}\,,  \nn\\
  B_{\Phi} &= -B_0 \cos{\left(\Phi + \psi
      \ln{\frac{\rho}{\rho_0}}\right)}  \cos{p} \cos{\chi}\,, \nn\\
  B_{z} &= B_0 \sin{\chi}\,.
\end{align}
Here, $\rho$, $\Phi$ and $z$ are Galacto-centric cylindrical
co-ordinates with $\Phi$, the cylindrical longitude, measured from the direction of the Sun, $p$
is the pitch angle of the field, $\psi = 1/\tan{p}$, $\rho_0$ defines
the radial scale of the spiral, $\chi = \chi_0 \tanh(z/z_0)$ parametrizes the
amplitude of the $z$ component and $z_0=1$~kpc. We use the parameters constrained in
\citet{2008A&A...490.1093M}: $p = -8.5$\,degrees, $\rho_0 = 11$~kpc
and $\chi_0 = 8$\,degrees, with the field amplitude set to $B_0 = 3
\mu$G, and take the distance between the Sun and the Galactic center
to be $8$~kpc. A diagram of the magnetic field orientation and magnitude in the BSS
model is shown in the left panel of Figure~\ref{fig:lsmodel}.

A number of other magnetic field models have been proposed in the
literature. For comparison we also include the Logarithmic Spiral Arm
(LSA) model introduced by \cite{2007ApJS..170..335P} for use in
cleaning of the \WMAP\ data.  The model is defined as
\begin{align}
  B_{\rho} &= -B_0 \sin{\left(\psi_0 + \psi_1
      \ln{\frac{\rho}{\rho_{\rm W}}}\right)}\cos{\chi}\,,  \nn\\
  B_{\Phi} &= -B_0 \cos{\left(\psi_0 + \psi_1
      \ln{\frac{\rho}{\rho_{\rm W}}}\right)}\cos{\chi}\,, \nn\\
  B_{z} &= B_0 \sin{\chi}\,,
\label{eq:chapthree9}
\end{align}
with parameters obtained by fits to the \WMAP\ K-band field directions;
$\psi_0=27$ degrees, $\psi_1=0.9$ degrees, and $\chi$ defined as in
the BSS model but with $\chi_0=25$ degrees. The radial scale is also
different in this model with $\rho_{\rm W}=8$~kpc whereas the scale
height is the same as above with $z_0=1$~kpc. There is no azimuthal
dependence in this model. The right panel of Figure~\ref{fig:lsmodel}
shows a slice through the Galactic plane for the LSA model.

Although both fields are unlikely to provide a full
description of our Galaxy \citep{2008A&A...486..819M,
  2008A&A...477..573S}, they are sufficient for our current purpose as we
do not require a precise template of the sky, only a reasonable
approximation against which to test foreground separation techniques
and the performance of experiments in the presence of systematic effects.

Both magnetic field models assume the field strength $B_0$ is constant
although there is weak evidence for some radial dependence
\citep{2006ApJ...642..868H}. Any such dependence will not affect the
polarisation model significantly and the overall radial dependence of
the signal is determined by the exponential drop-off in the dust
density which modulates the integrand along the line-of-sight. Field
reversals may also be present in the spiral arms but, if sharp enough,
will not contribute to the signal significantly.

\subsubsection{Small-Scale Galactic Magnetic Field}

The turbulent field is somewhat less well understood. When
constraining the above large-scale field, \citet{2008A&A...490.1093M}
simultaneously fit a small-scale field with best-fit r.m.s\ amplitude
$B_{\mathrm{r.m.s.}}  = 1.7 \mu$G. Several different studies agree
that the r.m.s.\ amplitude is similar to the amplitude of the
large-scale field in the Solar vicinity \citep{2002ApJ...564..762F,
  2006ApJ...642..868H}, and so here we set $B_{\mathrm{r.m.s.}} = 2
\mu$G.  \citet{1996ApJ...458..194M} examined the rotation measures of
extra-Galactic sources across a small patch of sky and concluded that
the data were consistent with Kolmogorov turbulence on scales smaller
than $4$\,pc, assuming a statistically isotropic, homogeneous Gaussian
field. On larger scales they found a somewhat flatter energy spectrum
with an outer scale of up to $96$\,pc. Kolmogorov-type spectra up to
kilo-parsec scales in the interstellar magnetic field and other
interstellar plasma components have also been reported by other
studies \citep{1995ApJ...443..209A, 2000ApJ...537..720L,
  2008arXiv0812.2023C}.

Kolmogorov turbulence describes the energy distribution among vortices
of different size, with the amplitude of the turbulence related to the
energy density at that position. Turbulent flow can be viewed as an
energy cascade from larger to smaller eddies. At small enough length
scales, known as the Kolmogorov length scale, energy is dissipated
through viscous dissipation. A Kolmogorov spectrum is proportional to
the rate of energy dissipation and the magnitude of the wavevector
$k$. Using the Kolmogorov energy spectrum one finds that the power
spectrum of a turbulent field is $\mathcal{P}(k) \propto
k^{-(2+3\,N_d)/3}$ where $N_d$ is the number of spatial dimensions of
the realisation. In this work, we generate realisations of this power
spectrum in order to model the three-dimensional magnetic field in
real space by a Fast Fourier Transform (FFT). 

It is numerically intractable to generate a realisation of this turbulent
field in three dimensions at sufficiently high resolution and to
accommodate the entire sky, hence we resort to independent
one-dimensional realisations along the line-of-sight to each
pixel. This model ignores correlations across the sky, but properly
incorporates the line-of-sight depolarisation. We choose an injection
scale of $100$~pc, assume the dissipation scale is small and use the
one-dimensional Kolmogorov energy spectral index of $-5/3$.

For smaller patches of the sky, relevant for ground-based
observations, a full, three dimensional realisation is feasible
together with a higher angular resolution in the line-of-sight
integrals. 

\begin{figure*}
\begin{center}\begin{tabular}{c}
     \makebox[3in][c]{\includegraphics[height=5cm, width=7cm,angle=180,trim=0cm 1cm 0cm 5cm,clip]{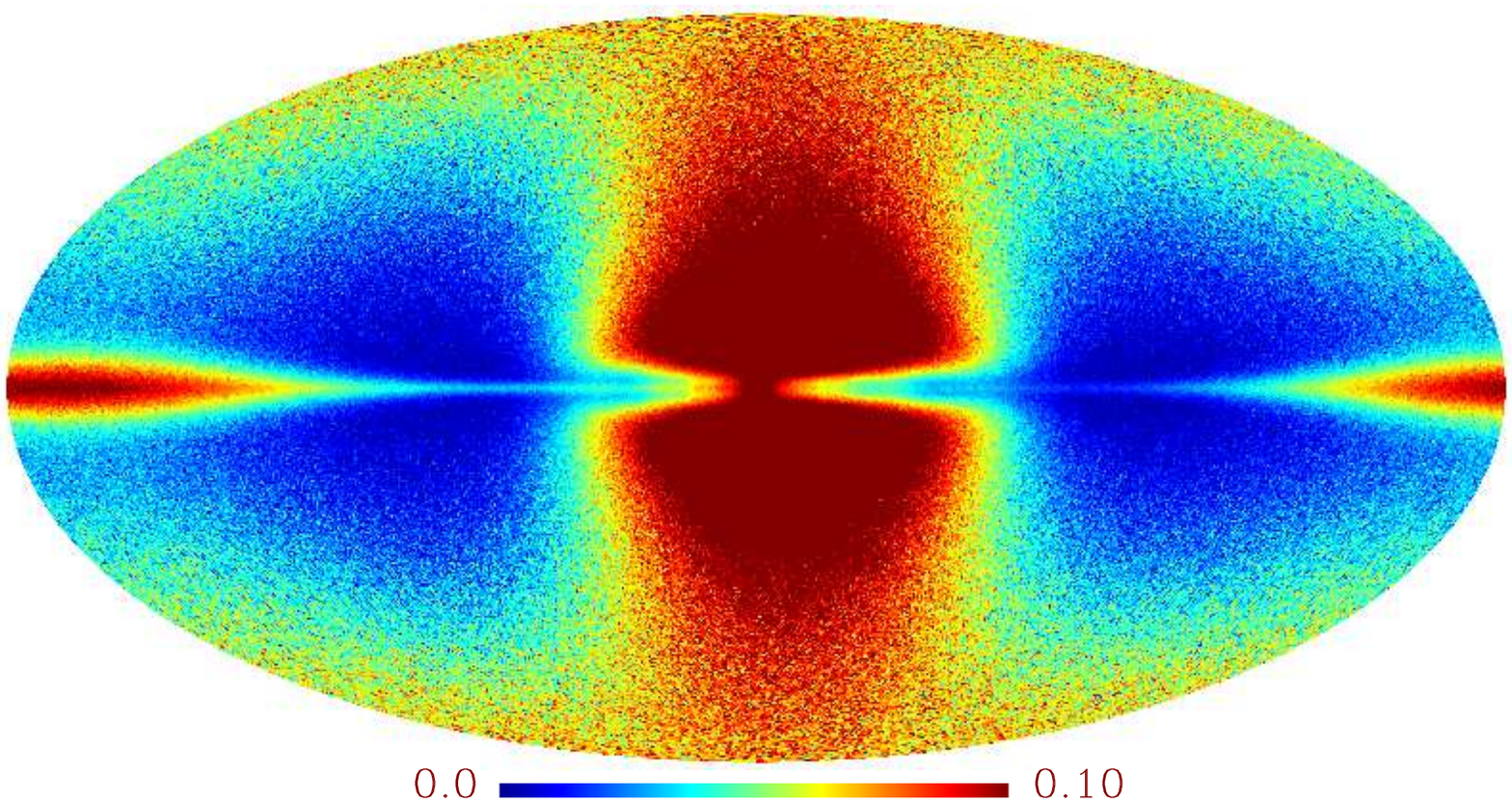} \includegraphics[height=5cm, width=7cm,angle=180,trim=0cm 1cm 0cm 5cm,clip]{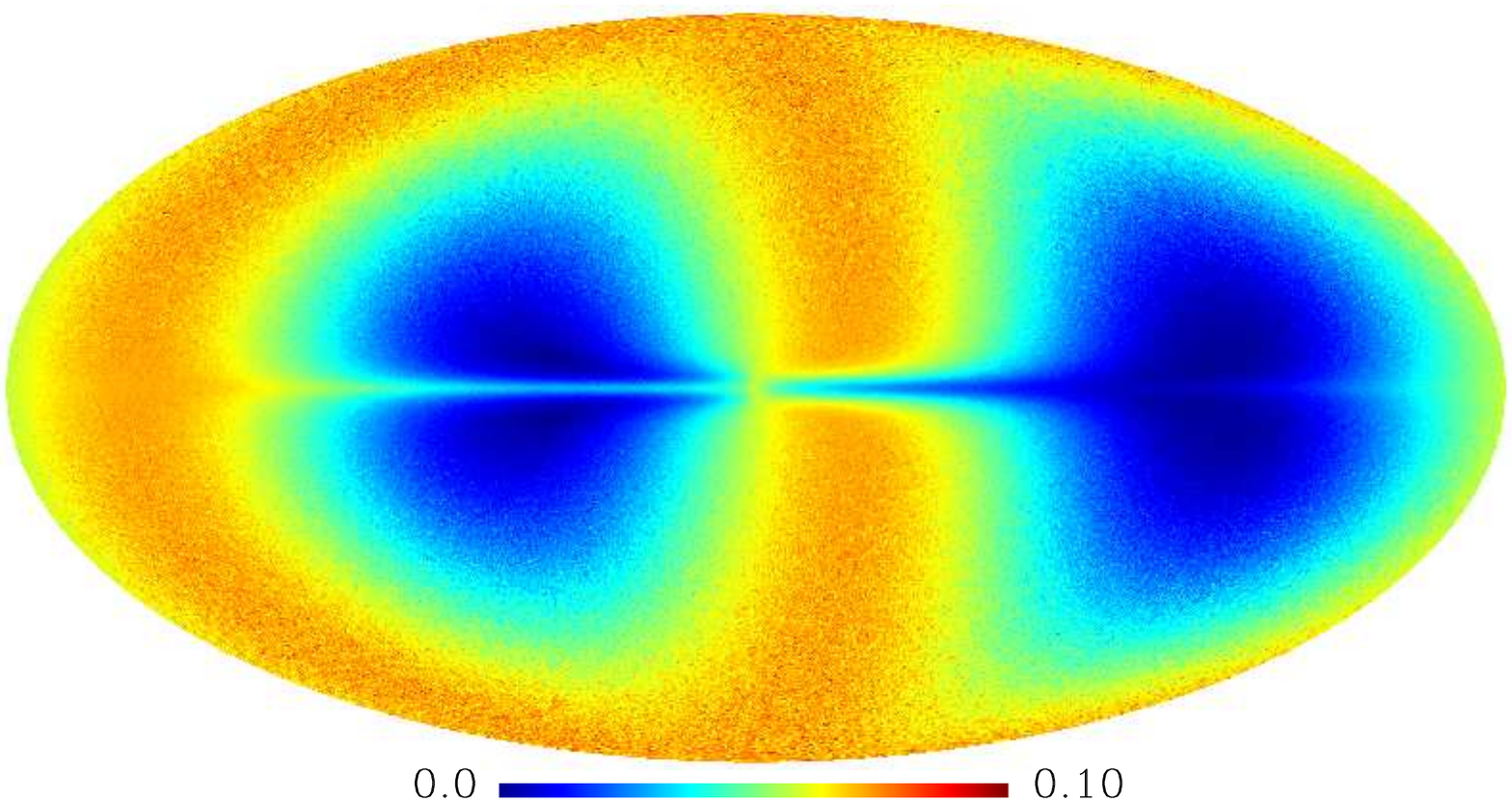}} \\
     \makebox[3in][c]{\includegraphics[height=5cm, width=7cm,angle=180,trim=0cm 1cm 0cm 5cm,clip]{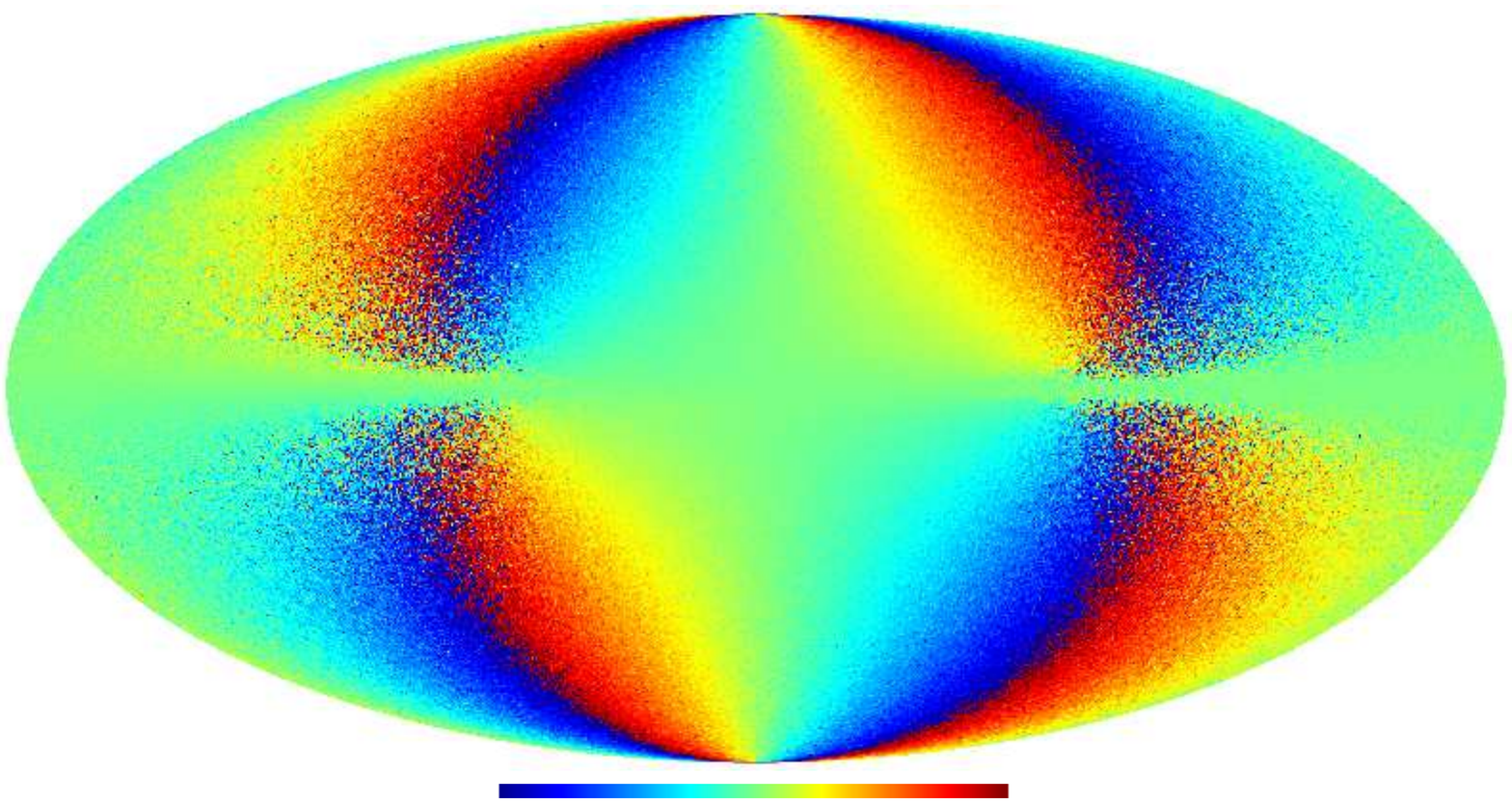} \includegraphics[height=5cm, width=7cm,angle=180,trim=0cm 1cm 0cm 5cm,clip]{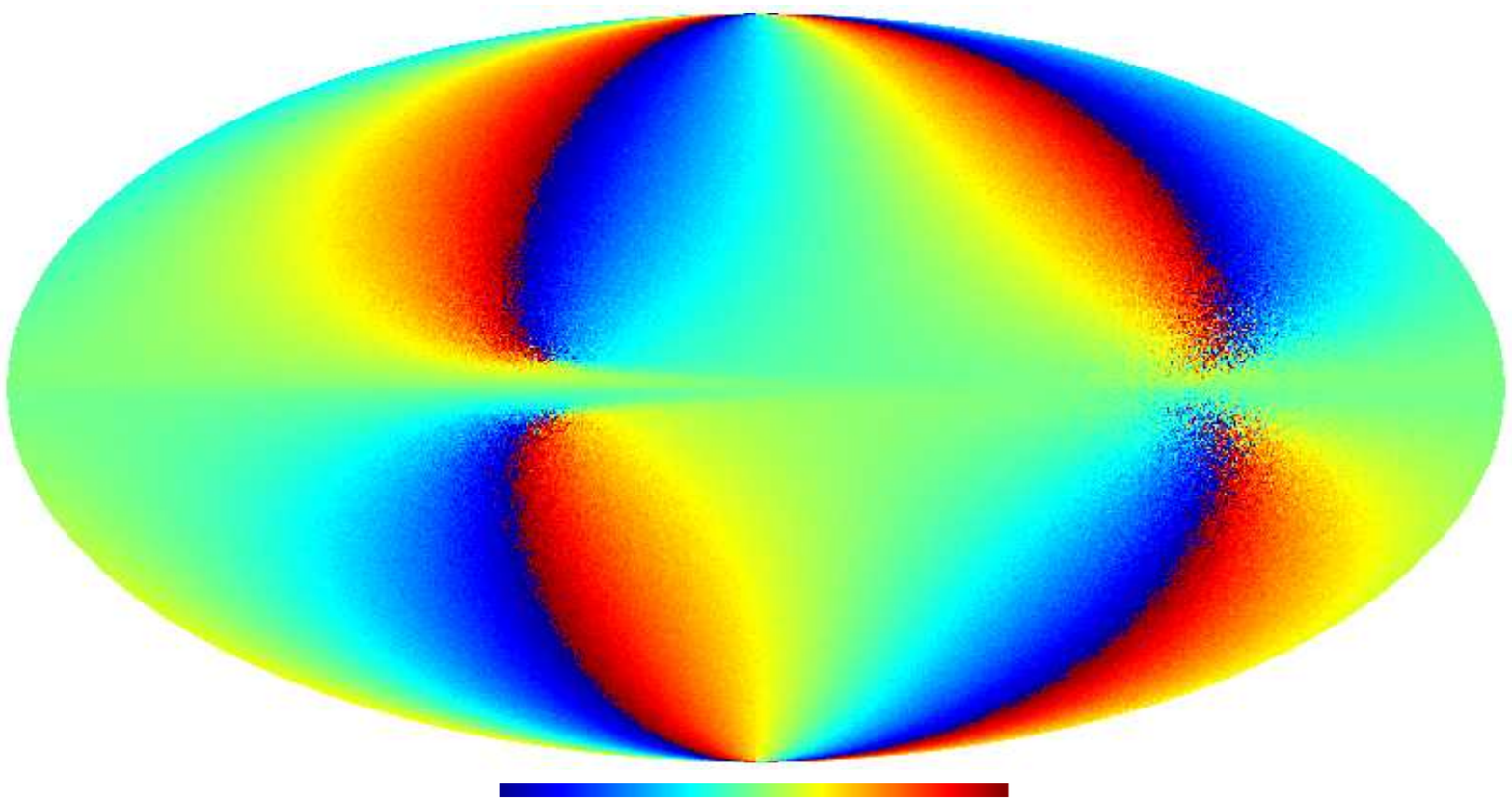}} \\
\end{tabular}
\caption[]{Polarisation fraction (top) and angle (bottom) in Galactic
  co-ordinates for our model of thermal dust emission at $150$\,GHz
  for the BSS (left column) and LSA (right column). The polarisation
  angle colour bar ranges from $-\pi/2$ to $\pi/2$. Both models
  include large (\lscale) and small (\sscale) scale magnetic field components. The \sscale\
  turbulent component was added in the one dimensional, line-of-sight
  approximation and can be seen as an uncorrelated noise addition to
  the coherent \lscale\ component. There are significant differences in the
  morphology of the polarisation fraction between the BSS and LSA
  models due to the BSS model including the spiral arm structure.}
\label{fig:pol}
\end{center}
\end{figure*}

\subsection{Dust Properties}
\label{sec:dustdens}
We model the large-scale spatial distribution of the dust density,
$n_d$, using a simplification of the model constrained in
\citet{2001ApJ...556..181D},
\begin{equation}
n_d = n_0 \exp{\left(-\frac{\rho}{\rho_d}\right)}
\sech^2{\left(\frac{z}{z_d}\right)}\,.
\label{eq:chapthree7}
\end{equation}

For consistency with the \WMAP\ polarisation analysis
\citep{2007ApJS..170..335P}, we take the scale height $z_d = 200$~pc
and the scale radius $\rho_d = 3$~kpc. We do not attempt to model the
small-scale variations in the dust density and temperature here, which
may also affect the polarisation degree and direction. Small-scale
variations in the total intensity are included via the FDS model.

The model also requires a description of the physics of grain
alignment and of the intrinsic polarisation of the emission from an
individual grain. In general these are complex functions of the
magnetic field and various properties of the grains. Recently, good
progress has been made in describing the details of the alignment
using the theory of radiative torques \citep{2007MNRAS.378..910L,
  2008MNRAS.388..117H}. However, it is still difficult to produce a
well-constrained quantitative description to apply to our model
\citep{2009arXiv0901.0146L}.

Instead, we describe the alignment in an integrated manner, without
recourse to the details of a particular physical mechanism. We assume
that the polarisation direction is always perpendicular to the
component of the magnetic field in the plane of the sky, and that the
degree of polarisation depends quadratically on the magnetic field
strength. This is similar to the behaviour assumed in
\citet{2007ApJS..170..335P}. We follow this approach in providing our
templates and do not attempt to account for any possible misalignment
of the axis of orientation of the dust grains with the magnetic field
lines, as is done in other work, for example
\citet{AA..526..A145..2011}.

\begin{figure*}
\begin{center}\begin{tabular}{c}
     \makebox[3in][c]{\includegraphics[height=5cm, width=7cm,angle=180,trim=0cm 1cm 0cm 5cm,clip]{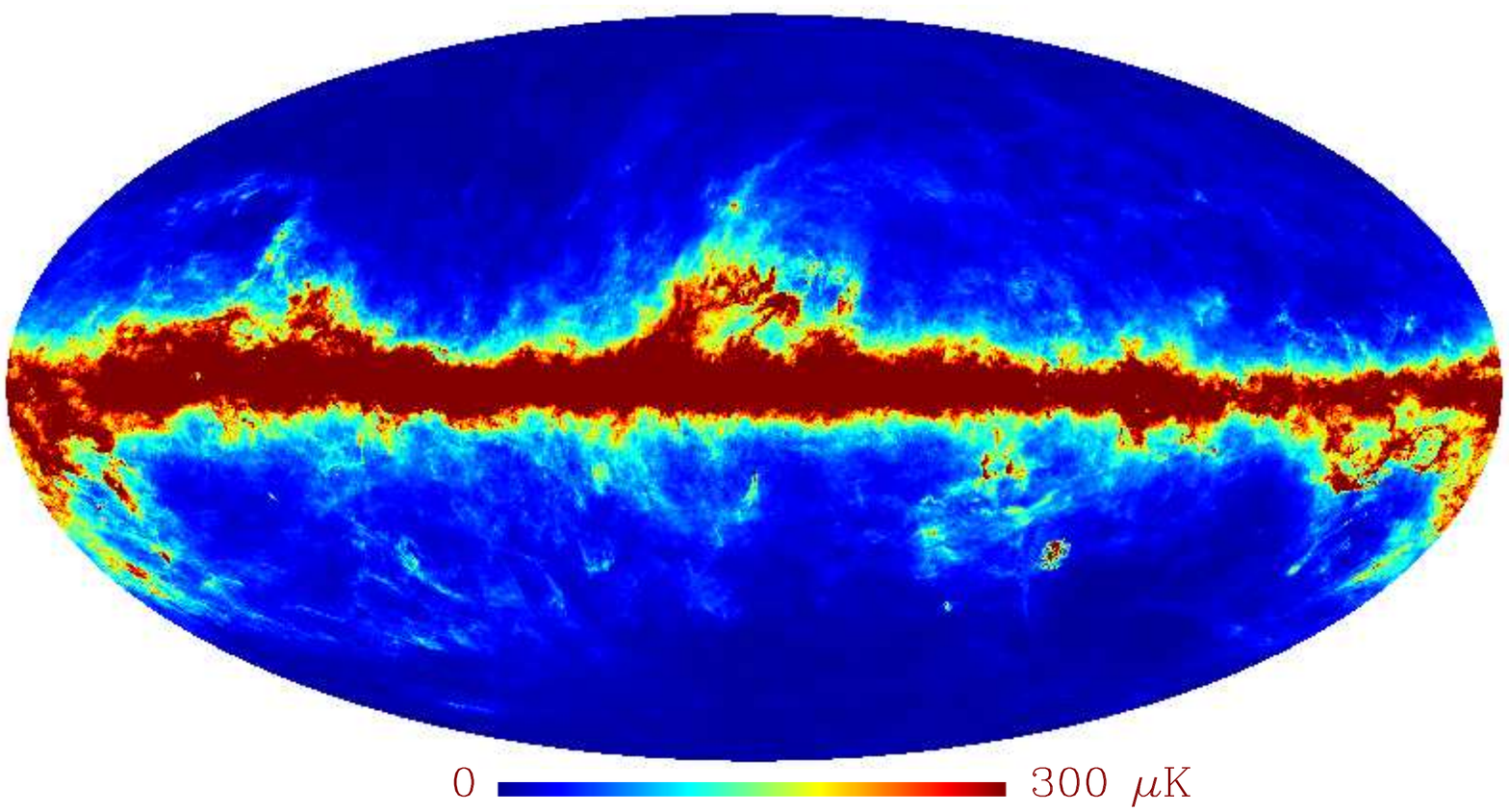} \includegraphics[height=5cm, width=7cm,angle=180,trim=0cm 1cm 0cm 5cm,clip]{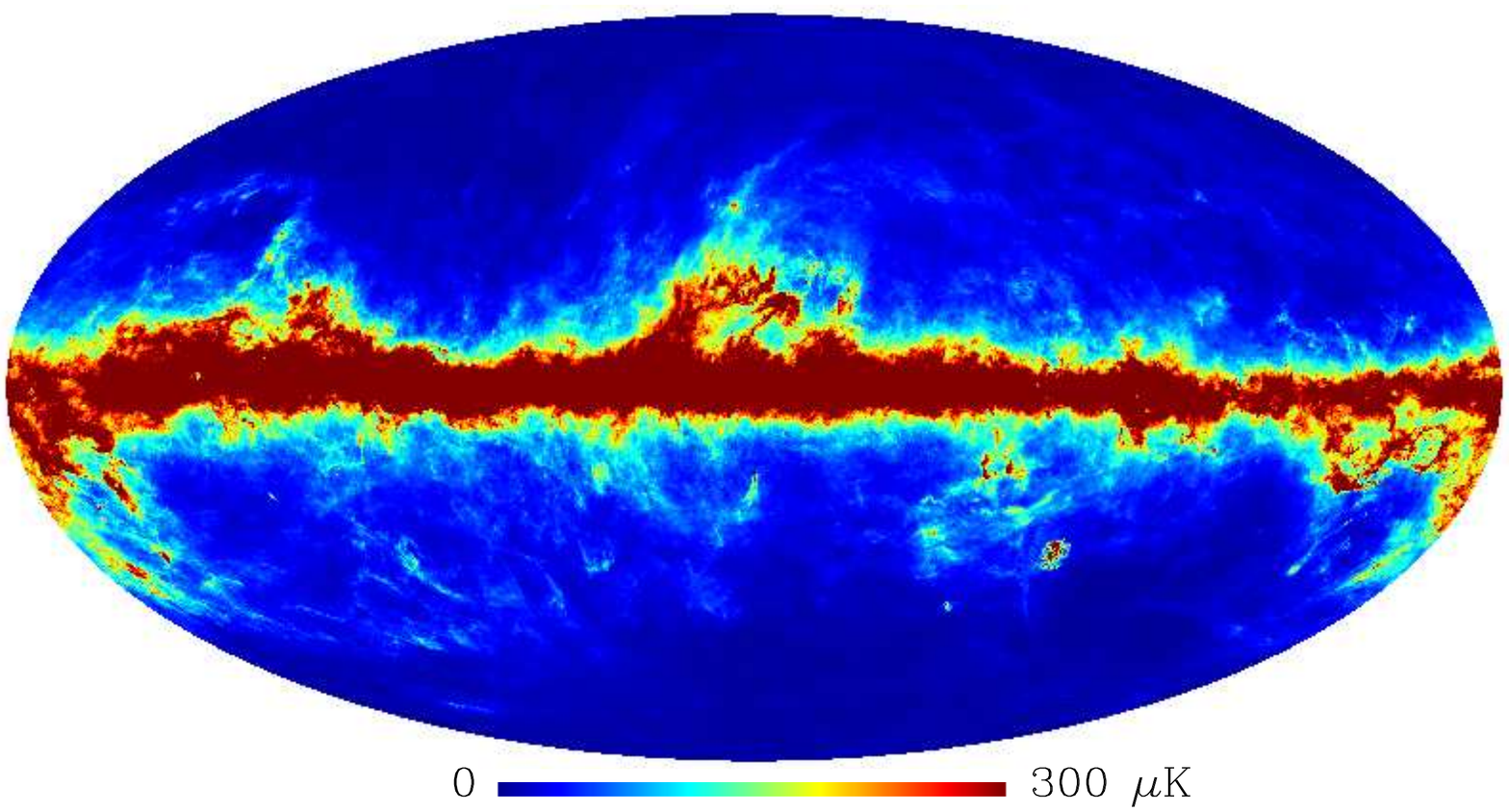}} \\
     \makebox[3in][c]{\includegraphics[height=5cm, width=7cm,angle=180,trim=0cm 1cm 0cm 5cm,clip]{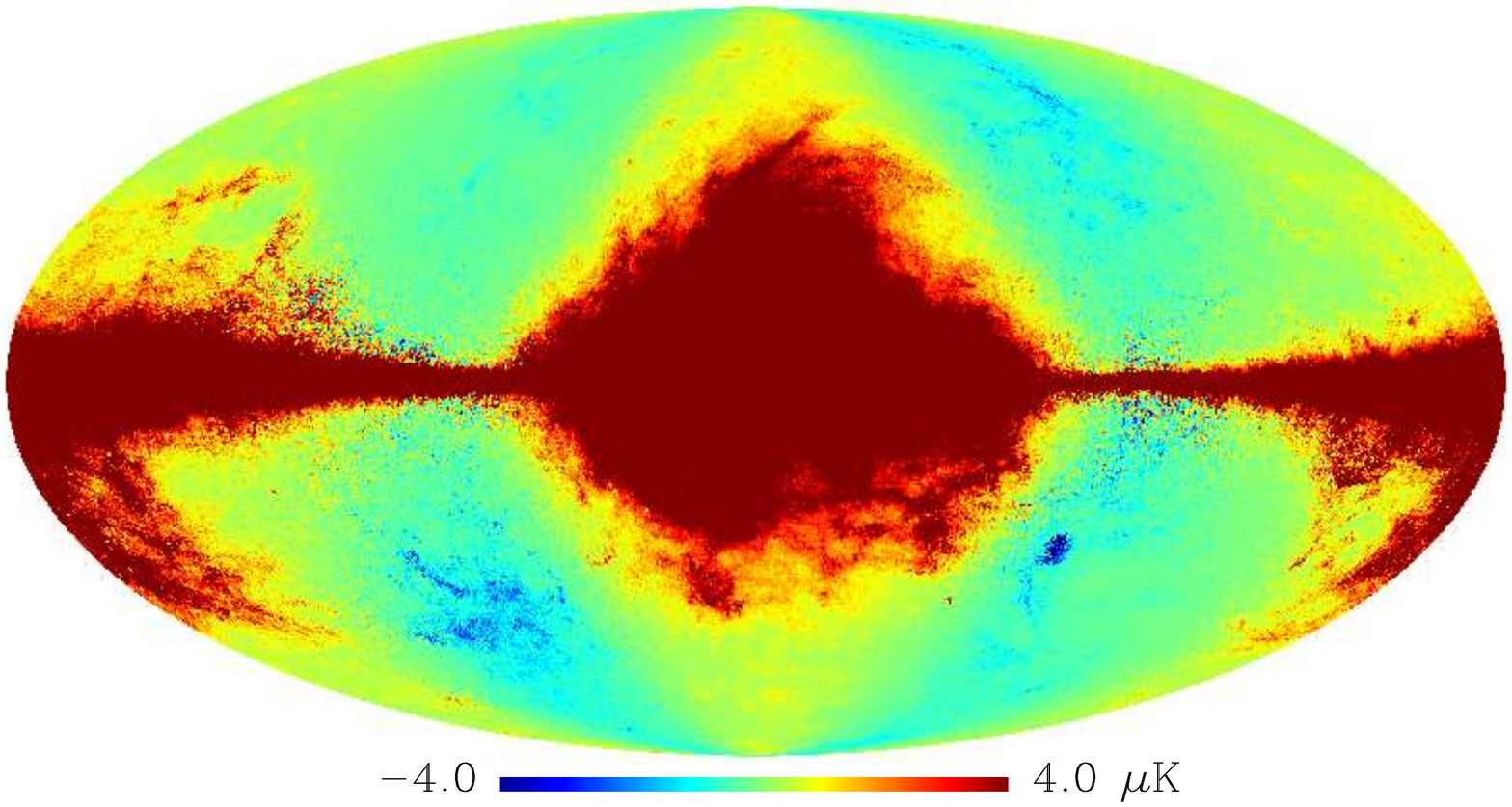} \includegraphics[height=5cm, width=7cm,angle=180,trim=0cm 1cm 0cm 5cm,clip]{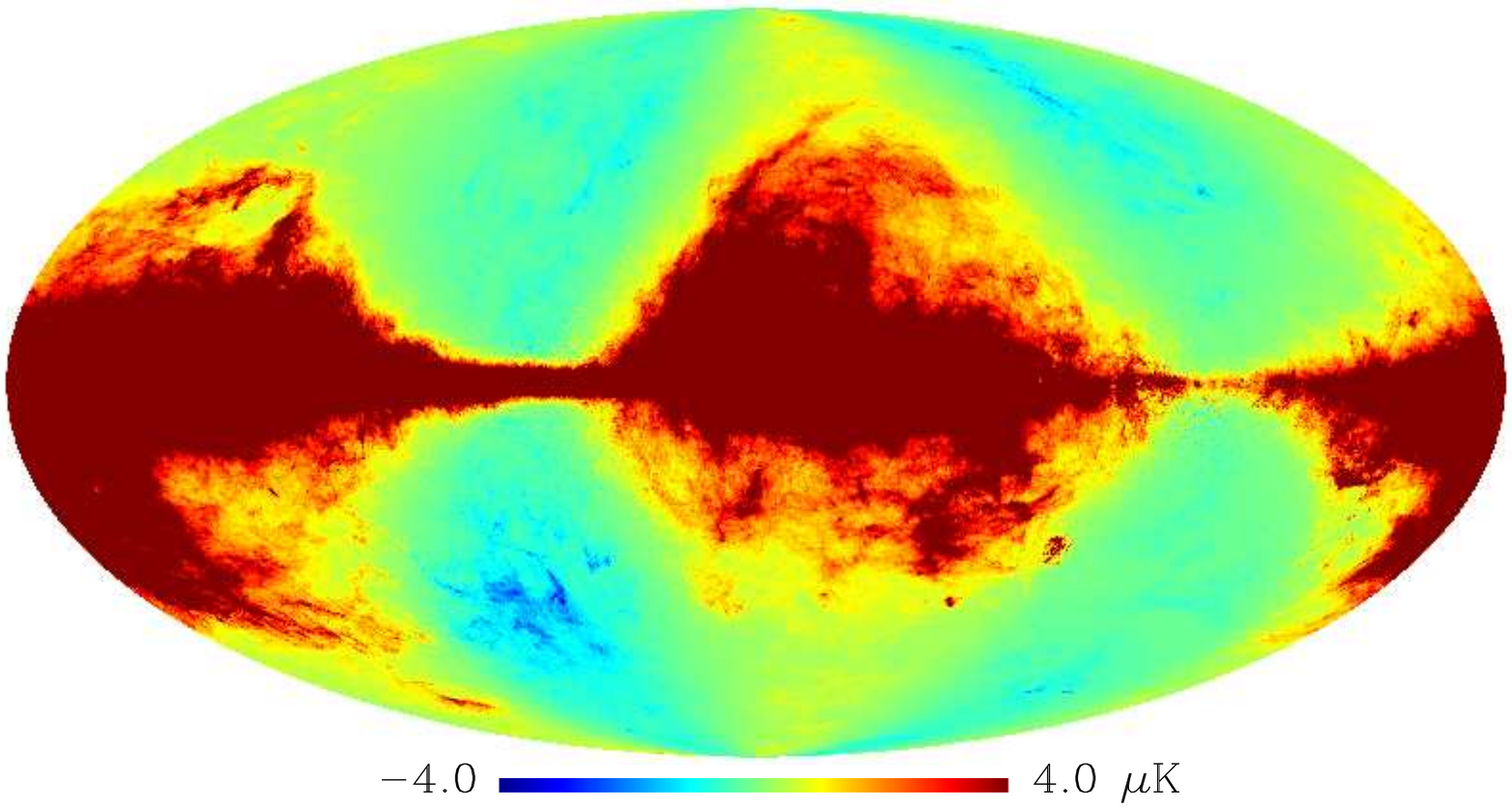}} \\
\makebox[3in][c]{\includegraphics[height=5cm, width=7cm,angle=180,trim=0cm 1cm 0cm 5cm,clip]{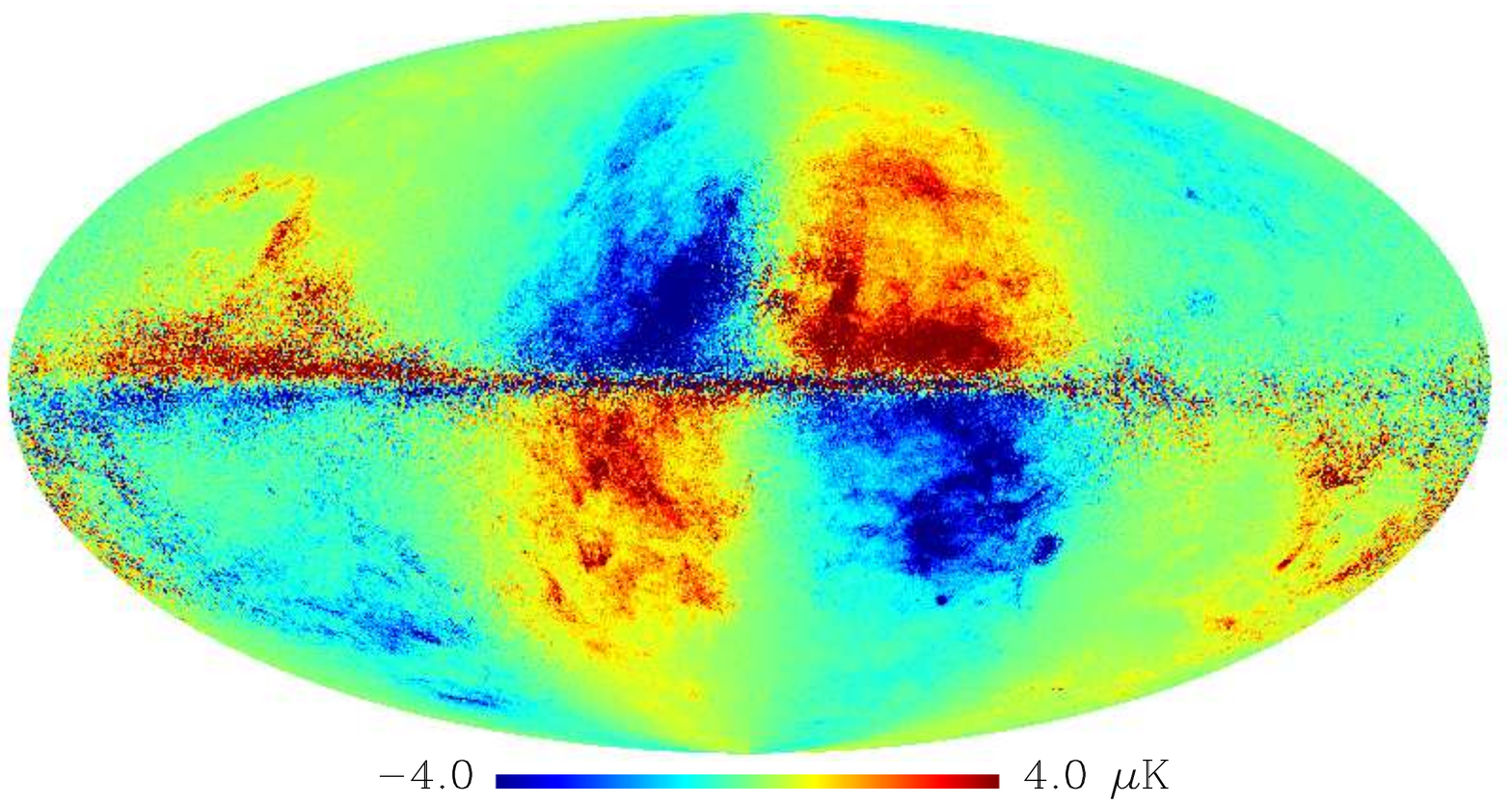} \includegraphics[height=5cm, width=7cm,angle=180,trim=0cm 1cm 0cm 5cm,clip]{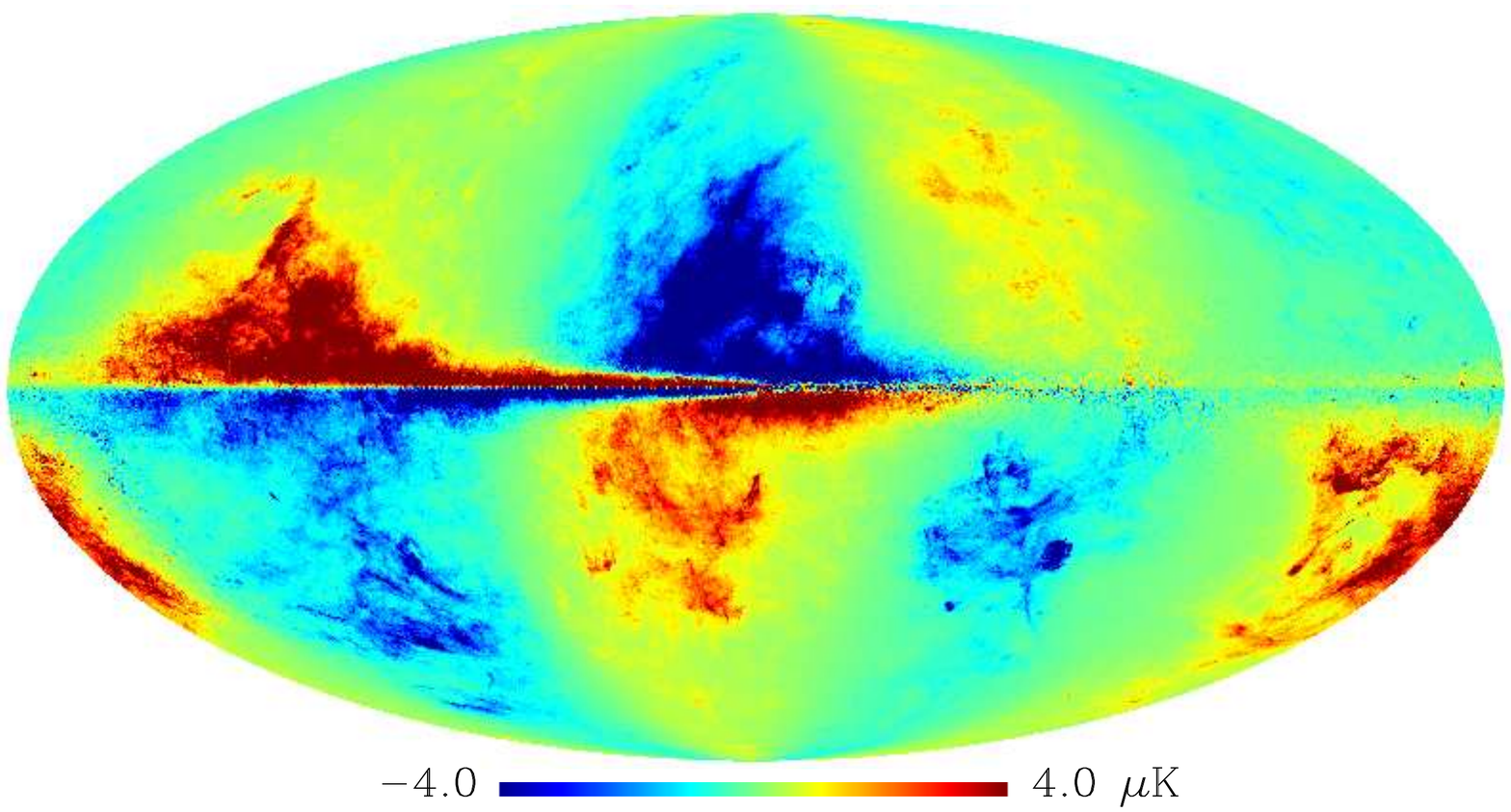}} \\
\end{tabular}
\caption[]{Stokes parameter maps (from top to bottom $I$, $Q$ and $U$)
  in Galactic co-ordinates for our model of thermal dust emission at
  $150$\,GHz for the BSS (left column) and LSA (right column).}
\label{fig:iqu}
\end{center}
\end{figure*}

\begin{figure*}
\begin{center}\begin{tabular}{cc}
     \includegraphics[scale=0.4]{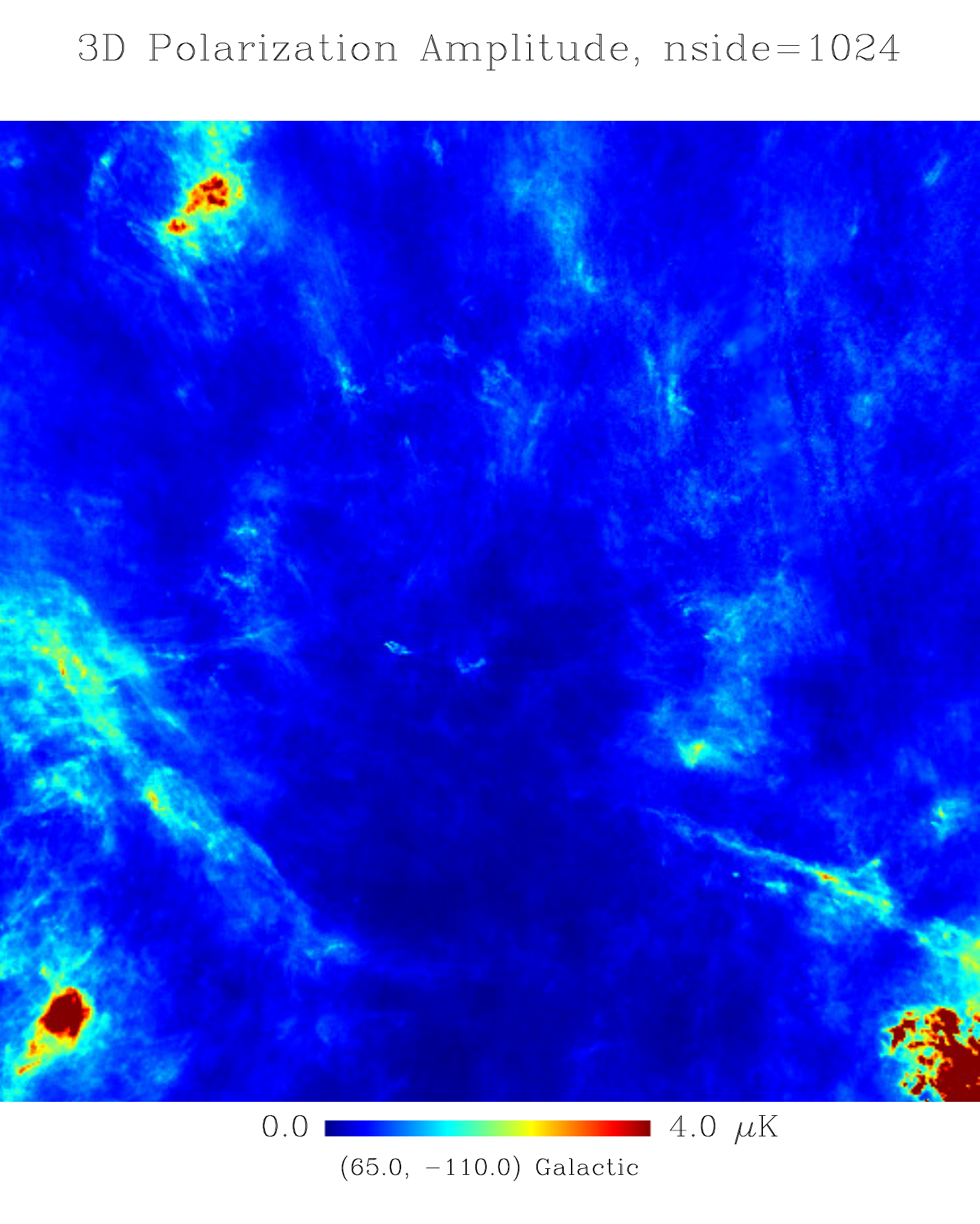}&\includegraphics[scale=0.4]{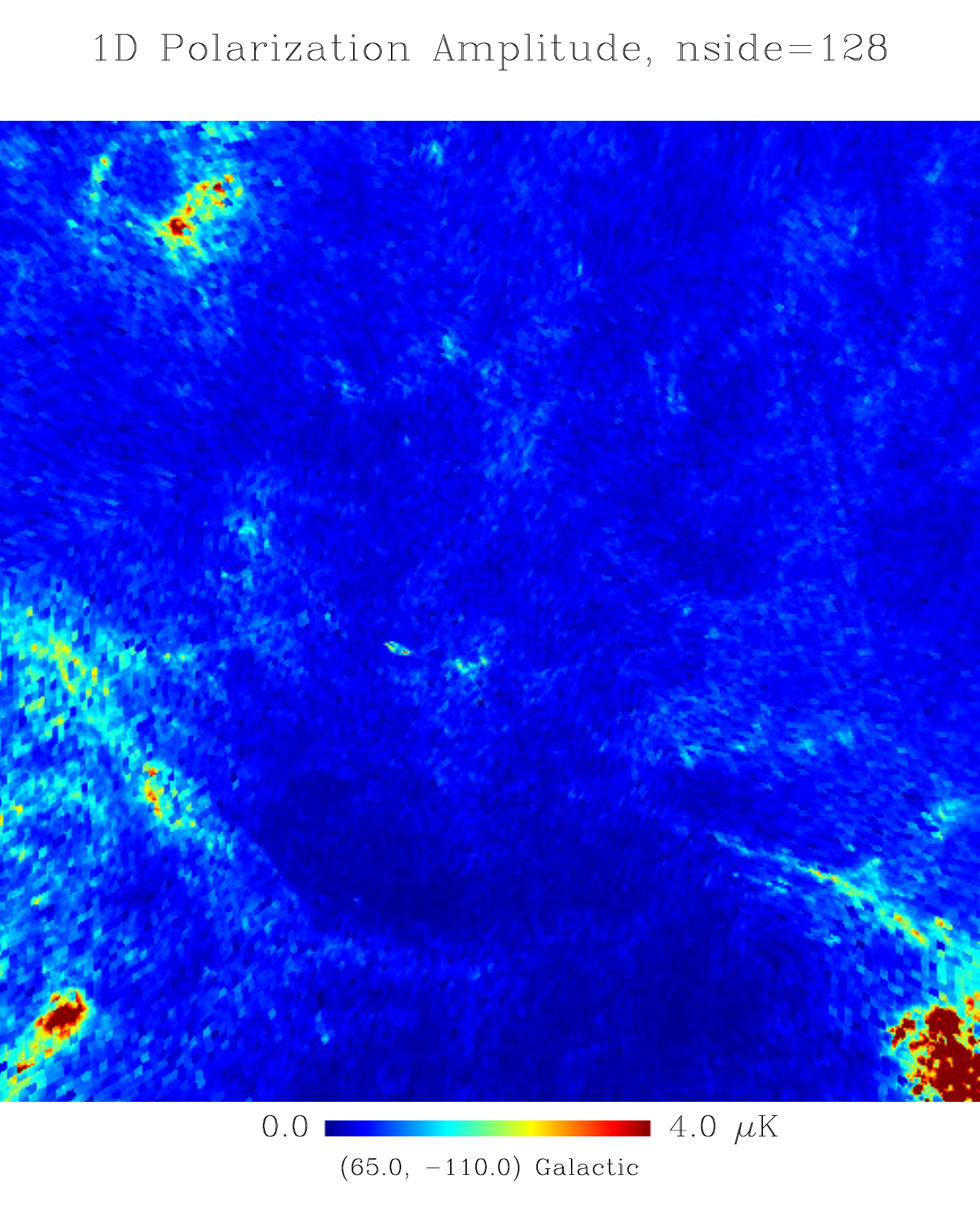}\\
     \includegraphics[scale=0.4]{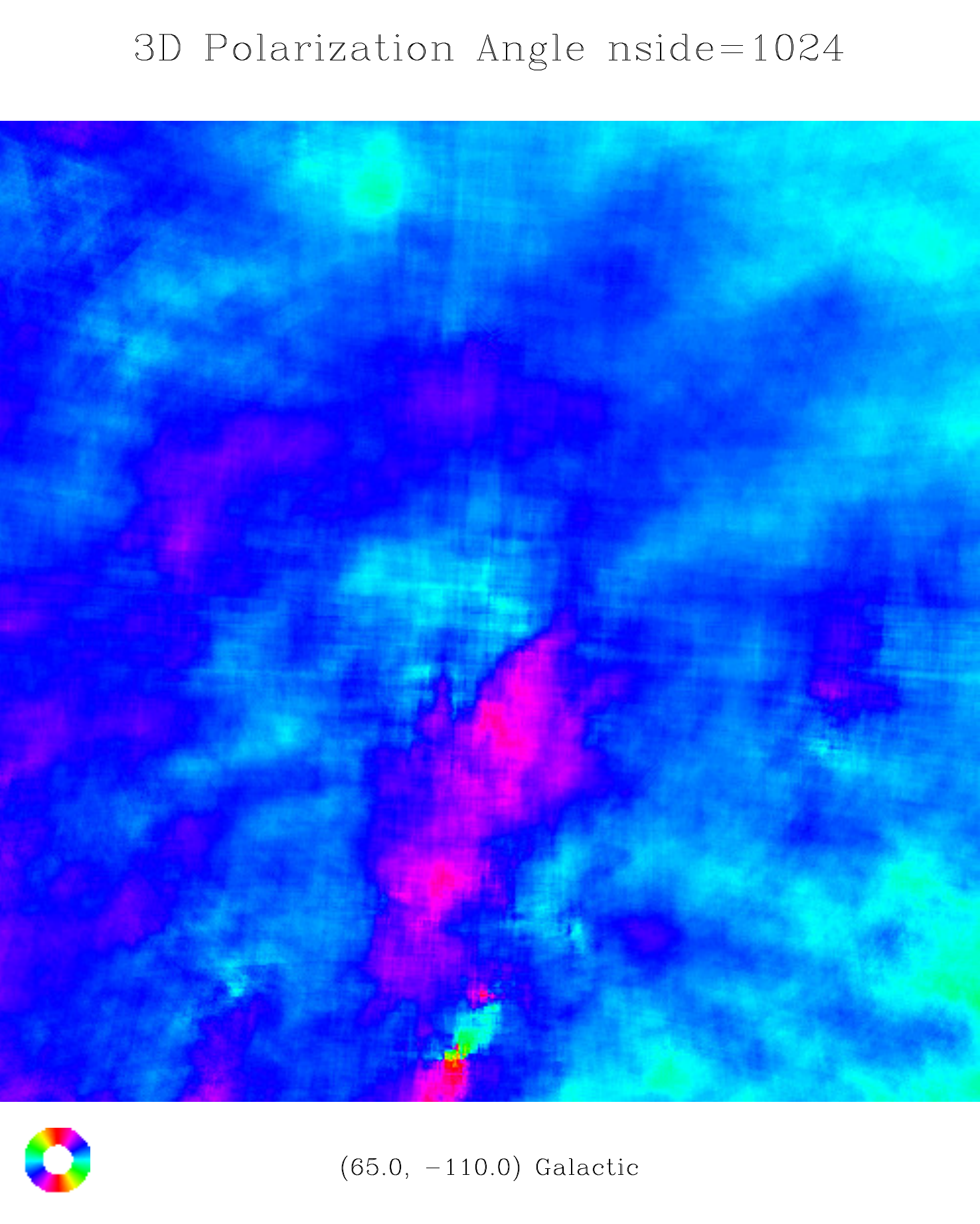}&\includegraphics[scale=0.4]{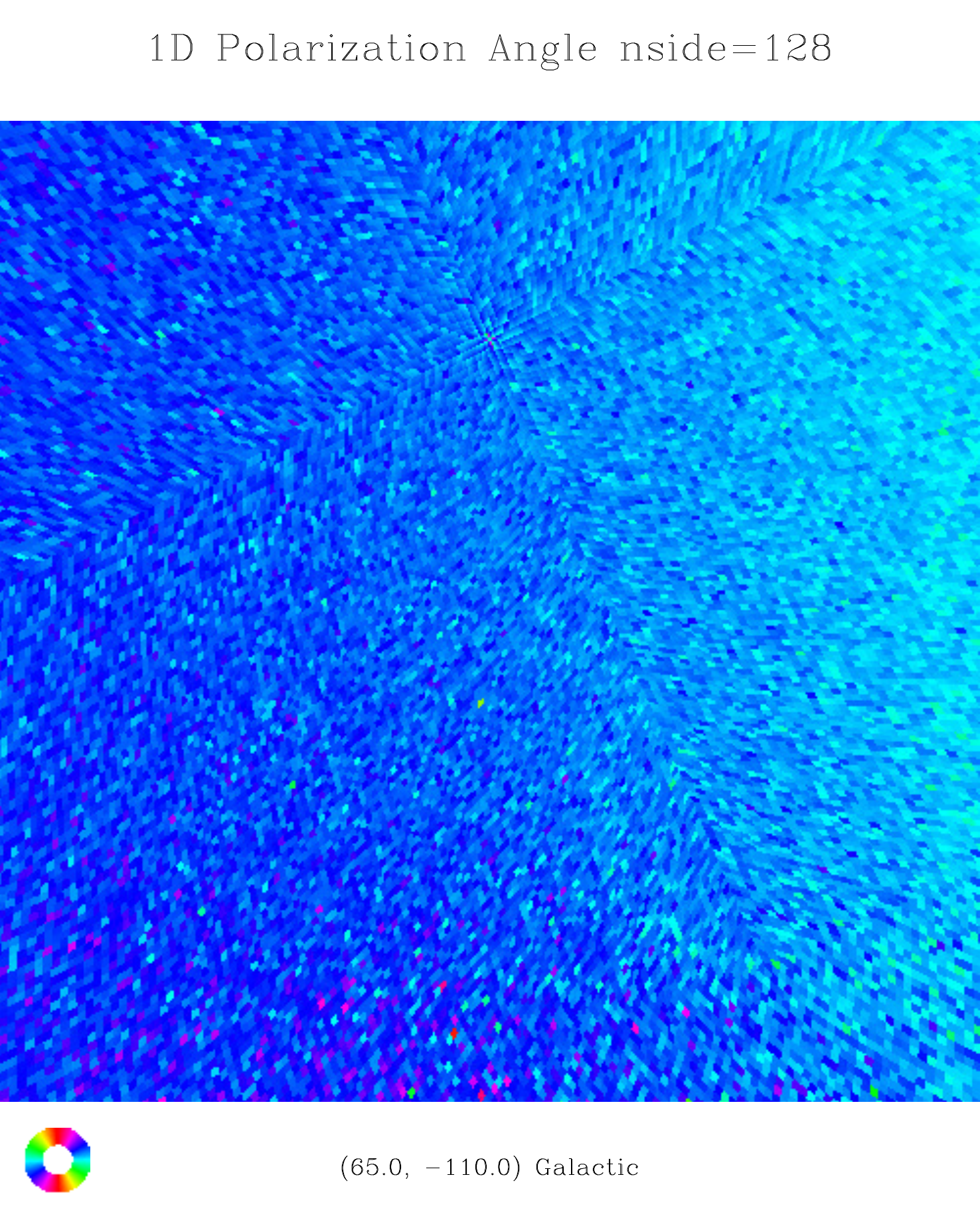}
\end{tabular}
\caption[]{Gnomic projection of the polarisation amplitude and angle
  in a $75 \times 75$ degree patch in the southern Galactic
  hemisphere. The left column shows the amplitude and angle from a
  line-of-sight integration including a full three dimensional
  realisation of the small-scale turbulent magnetic field
  model. $N_{\rm side}^{\rm P}=1024$ was used to calculate the ``3D''
  maps but only for lines-of-sight corresponding to pixels inside the
  patch. The right column shows the same area from the full-sky
  templates with $N_{\rm side}^{\rm P}=128$. The full-sky maps used a
  one dimensional realisation of the turbulent component along the
  line-of-sight to speed up the computation. The absence of correlated
  small-scale structure and lower angular resolution of polarisation
  information in the ``1D'' case is clearly seen when comparing maps.}
\label{fig:small}
\end{center}
\end{figure*}

\subsection{Stokes Parameters}
\label{sec:stokes}
We combine the small-scale (\sscale) and large-scale (\lscale) magnetic field
values according to \begin{align} B_r=B_{r,ss}+B_{r,ls}\,,\nn \\
  B_\theta=B_{\theta,ss}+B_{\theta,ls}\,, \nn \\
  B_\phi=B_{\phi,ss}+B_{\phi,ls}\,, 
\end{align}
where $r$, $\theta$, and $\phi$ are now Solar-centric spherical polar
co-ordinates. The polarisation at each point along the line-of-sight
$\hat r$ is determined by the perpendicular field components,
$B_\theta$ and $B_\phi$.

The Stokes parameters for this model are projected out from our
three-dimensional model using the appropriate line-of-sight integrals,
\begin{align}
  I_{\rm model}(\theta, \phi) &=\epsilon(\nu) \int_0^{r_{\max}} n_d(\vect{r}) \, \ud r\,, \nn\\
  Q_{\rm model}(\theta, \phi) &=\epsilon(\nu)\int_0^{r_{\max}} n_d(\vect{r}) p_0
  [B_\phi(\vect{r})^2 - B_\theta(\vect{r})^2] \, \ud r \,, \nn\\
  U_{\rm model}(\theta, \phi) &= \epsilon(\nu)\int_0^{r_{\max}} n_d(\vect{r}) p_0
  [2B_\phi(\vect{r})B_\theta(\vect{r})]\, \ud r\,,
\label{}
\end{align}
and the normalisation $p_0$ is set to reproduce the average
polarisation fraction reported by \WMAP\ outside their P06 mask, $3.6\%$
\citep{2007ApJ...665..355K}. Here, $\epsilon$
is the emissivity of the dust as a function of frequency, $\nu$. Note
that we conform to the default convention applied in the 
\healpix\footnote{See http://healpix.jpl.nasa.gov} package
\citep{2005ApJ...622..759G} regarding the sign of $U$.

We have chosen the $3.6\%$ average polarisation fraction as a
reference value but the templates can be scaled to fit any other
preferred value based on more detailed knowledge of the polarisation
fraction in smaller patches of the sky. It is also useful to note that
since we rescale the $Q$ and $U$ components the {\sl overall}
normalisation of the magnetic field model becomes irrelevant. However,
the {\sl relative} contributions from the \lscale\ and \sscale\ components in the
field remains as a model parameter.  

For the line-of-sight integrals we integrate from zero out to a
maximum line-of-sight distance $r_{\max}$ of 30,000~pc.  The integrals
are discretised in steps of $0.1$~pc. The direction of the
lines-of-sight are chosen to coincide with the centre of all \healpix\
pixels at a given $N_{\rm side}^{\rm P}$, where $N_{\rm side}^{\rm P}$
is less than or equal to $N_{\rm side}$ of the total intensity
template FDS map.

From this model we require maps of the polarisation direction,
$\gamma$, and degree, $P$, which are given by
\begin{align}
  P(\theta, \phi) &= \frac{\left( Q_{\rm model}^2 +
      U_{\rm model}^2\right)^{\frac{1}{2}}} { I_{\rm model}}\,, \nn\\
  \gamma(\theta, \phi) &= \frac{1}{2}
  \arctan{\left(\frac{U_{\rm model}}{Q_{\rm model}}\right)}\,.
\end{align}

Figure~\ref{fig:pol} shows maps of $P$ and $\gamma$ obtained from a
line-of-sight integration at resolution $N_{\rm side}^{\rm P}=128$ for the BSS
and LSA magnetic field models including a one dimensional, small-scale
turbulent component. The turbulent component is seen here as an
uncorrelated noise contribution to the large-scale correlations
induced by the large-scale magnetic field model. These maps can be
compared to the ``geometric suppression'' factor shown in the right
panel of Figure~8 of \cite{2007ApJS..170..335P}. There are significant
differences between the BSS and LSA field models in the morphology of
the polarisation fraction on the sky. The difference is greatest
towards the Galactic centre and bulge and the Galactic anti-centre
which coincides with a spiral arm. The LSA model does not include any
azimuthal dependence and as such does not model any modulation of the
magnetic field strength between spiral arms. In addition, the pitch
angle of the LSA model, as fit to the \WMAP\ data, is very low and this
leads to a very mild dependence of the field alignment in the radial
direction. These differences lead to a significantly simpler
polarisation structure in the LSA model than in the BSS case which
models the spiral arm structure explicitly.

The final dust model at frequency $\nu$ can be written as
\begin{align}
 I^\nu_{\dust} (\theta, \phi) &= I^\nu_{\fds}(\theta,\phi)\,,\nn\\
 Q^\nu_{\dust}(\theta, \phi) &= I^\nu_{\fds}(\theta,\phi)\,P(\theta,\phi) \,\cos\left(2\, \gamma(\theta,\phi)\right)\,, \nn\\
 U^\nu_{\dust}(\theta, \phi) &= I^\nu_{\fds}(\theta,\phi)\,P(\theta,\phi)\,\sin\left(2 \,\gamma(\theta,\phi)\right)\,,
\label{}
\end{align}
where $I_{\fds}^\nu$ is the total intensity FDS prediction. 

Our final product is a template foreground map with small-scale
structure modeled by the FDS predictions in the total intensity but
with polarisation fraction and angle determined {\sl internally} by
our magnetic field model and line-of-sight integrals. An alternative
approach taken by \cite{2007ApJS..170..335P} is to replace $\gamma$ with
a map $\gamma_{\dust} = \gamma_\star + \pi/2$ where $\gamma_\star$ is
a smoothed map of observed starlight polarisation directions. This
approach, however, is limited by the resolution of the starlight data
with only 1578 observations scattered around the sky. It also requires a
large smoothing kernel of approximately 10 degrees in size and limits
the application of any template derived in this way to very large
scales on the sky, corresponding to angular multipoles $\ell\lesssim
15$, and Galactic latitudes $\lvert b \rvert > 10\,^{\circ}$.

\section{Scales}
\label{sec:scales}
It is important to consider the range in angular scales our model is
valid for. All our maps are pixelised at $N_{\rm side}=1024$, this
ensures that the small-scale structure in the FDS prediction is
oversampled since the \IRAS\ resolution translates into a limit in
angular multipoles of roughly $\ell_{\fds}\sim 1700$ and the \healpix\
pixel smoothing scale is $\ell_{\rm pix}\sim 4\,N_{\rm side}$. The
overall, effective resolution of our templates is therefore limited by
the angular resolution of our line-of-sight coverage which is set by
the \healpix\ resolution $N_{\rm side}^{\rm P}$. 

For the full-sky maps presented here and made available publicly we
have chosen $N_{\rm side}^{\rm P}=128$ which corresponds to a limit of
roughly $\ell \sim 500$ in multipole space. We also show in our
example maps a small patch prediction with $N_{\rm side}^{\rm P}=1024$
(see Section~\ref{sec:maps}) which again oversamples the
  resolution given by the FDS templates.

It is also important to consider {\sl physically} relevant scales
that enable the interpretation of the structure in our templates. The
most important of these is the injection scale for the turbulent,
small-scale component of the magnetic field. We have set this to
100~pc. To obtain a rough estimate of the angular scales at which this physical
scale becomes important we can use a ``dust weighted'' distance
measure $\langle r\rangle = \int r\, n_d( \mathbf r)\, dr/\int n_d(
\mathbf r)\, dr\sim 7000$ pc for a mid-Galactic latitude
line-of-sight. This can be used to place the angular multipole scale
of injection at $\ell_{\rm inj}\sim 220$ or roughly 1 degree. Beyond
these scales the stochastic, turbulent component begins to
dominate the structure in the polarisation and the model is only a
statistical description of the real sky on these scales. An exhaustive 
exploration of foreground effects on scales below a degree would
therefore require a Monte Carlo approach.

\section{Maps}
\label{sec:maps}

We show a selection of template, full-sky maps at 150 GHz in
Figure~\ref{fig:iqu}. $I$, $Q$, and $U$ Stokes parameters are shown
for both BSS and LSA derived templates (other frequencies are
available on the on-line repository). The maps are of thermodynamic
CMB temperature in $\mu$K units and are shown in Galactic co-ordinates\footnote{Care must be taken in rotating Stokes parameters into
other co-ordinate systems such as ecliptic and we have provided rotated
maps on the on-line repository since most applications will simulate
observations in this frame.}. 

As detailed above, the $Q$ and $U$
components have been normalised such that the average polarisation
fraction outside the area defined by the \WMAP\ {\sl P06} mask is
3.6\%. The resolution of the \healpix\ maps is $N_{\rm side}=1024$ but
the polarisation information is based on a line-of-sight integral at
an angular resolution of $N_{\rm side}^{\rm P}=128$.

The maps have been obtained by the line-of-sight integration of a
magnetic field model that includes a small-scale turbulent realisation
only along the line-of-sight direction, ie. our ``one dimensional''
approximation. Whilst computationally intensive, ``3D'', full-sky maps
that include a full three dimensional realisation of the turbulent
component can be obtained, if required, with computation times of the
order of 10 days. However we show results for a smaller $75\times75$
degree patch in the southern Galactic hemisphere in
Figure~\ref{fig:small}. These maps were obtained using a full three
dimensional realisation at an angular line-of-sight resolution of
$N_{\rm side}^{\rm P}=1024$ and are compared with the same patch in
the full-sky ``1D'' maps. The difference between the two is most
clearly seen in comparing the polarisation angle which is uncorrelated
with the FDS intensity template. The full three dimensional case
contains correlated structure on smaller scales due to the coloured power
spectrum of the realisation. In contrast the one dimensional case is
uncorrelated on small scales whilst preserving the large-scale
correlations induced by the fixed, large-scale magnetic field
model. Tailored, high-resolution, ``3D'' realisations of small patches
such as those shown in Figure~\ref{fig:small} are most useful for sub-orbital 
experiments that can only observe a limited fraction of the
sky. 

\section{Conclusions}
\label{sec:conclude}
We have described a model of the polarised foreground that we expect to
observe due to emission from dust within our Galaxy. The model uses a
three-dimensional model of the Galactic magnetic field and dust field
and integrates along the line-of-sight to each \healpix\ pixel to
obtain a polarisation amplitude and angle. This information is
combined with total intensity, FDS derived template maps at different
frequencies to obtain a complete, polarisation template of foreground
emission by interstellar dust. 

We have concentrated on two popular models for the structure of the
large-scale structure of the Galactic magnetic field, namely, the BSS
and LSA models. The parameters for the BSS model have been calibrated
directly from measurements of the strength of the Galactic magnetic
field. In the LSA case we have employed the parameters obtained by
\cite{2007ApJS..170..335P} in fitting to the \WMAP\ observations. We
calculate the polarisation alignment {\sl internally} to our model in
both cases since there is not sufficient external information on
polarisation angles at resolutions relevant in this work. Some
differences exist between the BSS and LSA derived templates but these
are mostly at low Galactic latitudes away from the Galactic centre and
as such experiments targeting small areas at high Galactic latitudes
will not be sensitive to the differences. The differences do indicate
however that a more accurate model of the Galactic magnetic field is
required to produce realistic polarisation templates for low Galactic
latitudes.  In the future, the \planck\ mission will provide an important test of Galactic magnetic field models through detailed characteristion of galactic foregrounds.

We have developed a one dimensional approximation of the stochastic,
turbulent, small-scale component of the field for obtaining full-sky
templates. A full three dimensional realisation of the turbulent
component can be used to obtain higher resolution templates for
smaller patches of the sky. 

In future work we will be extending the model to include synchrotron
emission to form a complete picture of foreground emission relevant
for polarisation experiments. Other developments will be required to
increase the fidelity of the templates on small scales. These include
the addition of a stochastic, small-scale density field to model small-scale 
structure in the density. In full ``3D'' calculations this will
require the generation of an additional three dimensional, turbulence
realisation which is correlated to the small-scale magnetic
field. In addition, it would be useful to develop a simple model for
the correlation of both realisations with the small-scale structure in
the FDS derived total intensity templates.

There is significant freedom in the parameters defining the small-scale 
structure in the templates. Experiments targeting small angular
scales over small patches of the sky will be most sensitive to
variations in the parameters and also to the stochasticity of the
structure in the templates. Further Monte Carlo explorations of the
variation in the maps is therefore warranted to quantify the impact of
foregrounds on future sub-orbital experiments. As part of future work
we will generate large ensembles of random realisations of the
templates on small patches of the sky for the purpose of Monte Carlo
studies. 

The maps obtained from this model are available for use and can be
downloaded from an on-line repository\footnote{\url{http://www.imperial.ac.uk/people/c.contaldi/fgpol}}.

\section*{Acknowledgments}
We acknowledge useful discussions with the \spider\
team and in particular with Cynthia Chiang and Aur\'{e}lien
Fraisse. Daniel O'Dea and Carlo Contaldi acknowledge support from
STFC under the standard grant scheme (PP/E002129). Caroline Clark is
supported by an STFC studentship. Carolyn MacTavish is supported by a
Kavli Institute Fellowship at the University of Cambridge.
Calculations were carried out on a facility provided by the Imperial
College High Performance Computing
Service\footnote{\url{http://www.imperial.ac.uk/ict/services/teachingandresearchservices/highperformancecomputing}}.

\bibliography{ForegroundsBib}

\label{lastpage}

\end{document}